\documentclass[reprint,superscriptaddress,amsmath,amssymb,aps,prl]{revtex4-1}
\pdfoutput=1
\usepackage[utf8]{inputenc}
\usepackage{physics}
\usepackage{siunitx}
\usepackage{graphicx}
\graphicspath{{../Images/}}

\begin{document}
\title{Temperature Equilibration Due to Charge State Fluctuations in Dense Plasmas}

\newcommand{\Phys}
{Plasma Physics Group, Blackett Laboratory, Imperial College London, London, SW7 2AZ, UK}

\author{R.~A.~Baggott}
\email{r.baggott@imperial.ac.uk}
\affiliation{\Phys}

\author{S.~J.~Rose} 
\affiliation{\Phys}

\author{S.~P.~D.~Mangles} 
\affiliation{\Phys}

\begin{abstract}
The charge states of ions in dense plasmas fluctuate due to collisional ionization and recombination.
Here we show how, by modifying the ion interaction potential, these fluctuations can mediate energy exchange between the plasma electrons and ions.
Moreover, we develop a theory for this novel electron-ion energy transfer mechanism.  
Calculations using a random walk approach for the fluctuations suggest that the energy exchange rate from charge state fluctuations could be comparable to direct electron-ion collisions.
This mechanism is, however, predicted to exhibit a complex dependence on the temperature and ionization state of the plasma, which could contribute to our understanding of significant variation in experimental measurements of equilibration times.
\end{abstract}

\maketitle

Plasmas at high temperatures and densities are found throughout the Universe, including in stars, giant planets and supernovae \cite{Woosley2002,Chabrier2006,Fortney2009}.  
With the latest generation of experimental facilities, including high-power lasers \cite{Moses2009} and x-ray free electron lasers (XFELs) \cite{Glenzer2016}, it is becoming increasingly possible to explore these extreme conditions in the laboratory.  

Many of the mechanisms by which dense plasmas may be created impart energy predominantly to either the plasma electrons or the ions.  
Lasers or x-rays heat the electrons, while leaving the ions relatively cool.  
Shock heating meanwhile, creates a hot ion population while only weakly heating the electrons.  
This is also true in astrophysical shocks, where separate electron and ion temperatures exist behind the shock front \cite{Bykov2008,Rakowski2005}.  
Understanding how the electrons and ions exchange energy to achieve a common temperature is therefore key to understanding not only experiments, but also various astrophysical phenomena.  
Electron-ion energy equilibration is also important for inertial confinement fusion \cite{Xu2011}, where it facilitates the transfer of alpha-particle energy to the fuel ions, which is a prerequisite for ignition \cite{Hurricane2016}.

Despite its importance, the exchange of energy between electrons and ions remains poorly understood \cite{Ng2012,Hartley2015}.  
The earliest calculations, made by Landau \cite{Landau1936,*Landau1937} and Spitzer \cite{Spitzer1956}, considered classical, binary collisions between electrons and ions.  
This approach was subsequently extended to account for quantum effects and stronger interactions between the plasma particles \cite{Brysk1974,Lee1984}.  
These relatively simple Spitzer-type models achieve reasonable accuracy in weakly-coupled plasmas \cite{Gericke2002} and remain widely used for computational applications.

More recently, energy exchange has been described in terms of interactions between electron and ion density fluctuations \cite{Dharma-wardana1998}.  
This approach describes the same physical process as earlier work, and approximately reproduces the Landau-Spitzer result under appropriate conditions \cite{Hazak2001}.  
However, it is more readily generalized to include collective effects, including screening and coupled electron-ion modes \cite{Vorberger2009}, as well as correlations \cite{Daligault2009}, which become important for dense plasmas.  
Calculations including coupled collective modes \cite{Dharma-wardana1998,Vorberger2009} predicted lower rates than the Spitzer-based models.

The lower rates predicted by coupled-mode models seemed to be supported by experimental results showing long equilibration times in shock-heated material \cite{Celliers1992,Ng1995,Riley2000}.
However, subsequent measurements have shown much shorter timescales \cite{Hau-Riege2012,Hartley2015}.  
Measured equilibration times vary from sub-picosecond \cite{Hau-Riege2012} up to nanosecond \cite{White2014} timescales, even under conditions that seem superficially similar.
Experimental measurements of electron-ion equilibration remain challenging due to the need for accurate, time-resolved temperature diagnostics.
Ultrafast x-ray diagnostics are opening new avenues in this regard \cite{Zastrau2014,Mahieu2018,Kettle2019}.
Nonetheless, our understanding of energy exchange in these systems remains incomplete.

Electron-ion energy transfer, along with other properties such as the opacity \cite{Huebner2014} and the equation of state \cite{Das2009}, depends on the ionization state of the plasma.
In existing calculations, partial ionization can be accounted for most easily using the mean ionization, or more accurately using statistically averaged charge state distributions.
In local thermodynamic equilibrium, the charge state distribution can be obtained from the Saha equation.
On shorter timescales, the time-dependent ionization state must be obtained by solving rate equations incorporating the ionization and recombination rates \cite{Bornath1993}.
As the free electrons must supply the required ionization energy, these rate equations are coupled to the energy balance and so can indirectly influence electron-ion energy transfer \cite{Ohde1996,Bornath1998}.

The charge state distribution obtained by solving rate equations tends towards a steady state.
However, even in steady state, the charge state of individual ions will fluctuate due to ongoing ionization and recombination.
Because ion-ion interactions are screened by the bound electrons, any changes in charge state modify the interaction between an ion and its neighbours.
We suggest that, over a cycle of ionization and recombination, these changes in interaction strength can lead to a net transfer of energy between the ions and the electrons.
The basic physical process can be described as follows:
Upon ionization, the potential energy between a central ion and its neighbours is increased, with the additional energy supplied by the free electrons that drive collisional ionization.
The nearby ions are then repelled, gaining kinetic energy.
After recombination, the interaction strength is reduced, allowing the ions to reapproach the central ion whilst retaining some of the additional energy.
The process can also run in reverse, with the ions losing kinetic energy to work against the repulsion.

A similar mechanism is observed in the creation of ultracold plasmas, where changes in interaction strength upon initial ionization lead to a phenomenon known as disorder- or correlation-induced heating \cite{Gericke2003,Murillo2006}, and was suggested in an earlier work as a possible limitation on gain in soft x-ray lasers \cite{More1986}.
Other mechanisms involving charge fluctuations have also been studied, in dusty and space plasmas in particular \cite{Vaulina1999,Evans2008,Wang2009}.

In this work, we develop a theoretical framework to study energy transfer due to the charge state fluctuation mechanism.
Our approach is similar to that used to describe direct electron-ion interactions in terms of density fluctuations, except that we treat fluctuations in ion charge instead of electron density.
The energy transfer rate is then evaluated using a random walk model for the charge state fluctuations in some archetypal dense plasma systems.

The long-range interactions between plasma particles are most conveniently described in reciprocal space, i.e. as a function of frequency $\omega$, and wavevector $\vb{k}$.
When the charge state of an ion fluctuates by an amount $\var{Z}$, the change in the potential around the ion is given by
\begin{equation}
    \var{\phi_Z(\vb{k}}, \omega) = \frac{4 \pi e}{k^2} \var{ Z(\omega) },
\end{equation}
where $e$ is the electron charge and $k=\abs{\vb{k}}$.
Here, we have assumed that the bound electrons are strongly localized.
As any electrons that become significantly delocalized are considered ionized, this assumption should not be violated to any great extent, and could be relaxed in principle.
Within linear response theory, the change in potential induces a corresponding change in the surrounding ion density
\begin{equation}
    \var{ n_i(\vb{k}, \omega)} = \chi_{ii}(\vb{k}, \omega) \frac{4 \pi e}{k^2} \var{ Z(\omega) }
\end{equation}
where $\chi_{ii}$ is the response function of the surrounding ions \cite{Sturm1993,Ichimaru2004}.
The change in potential acts on the induced density change to transfer energy to the ions.
The rate at which energy is transferred to the surrounding ions is given by \cite{Ichimaru1975}
\begin{multline}
    \dv{E_i}{t}_{e \rightarrow i} = n_i \int \frac{\dd{\vb{k}}}{(2\pi)^3} \int \frac{\dd{\omega}}{2\pi} e \\
    \times \langle \var{\vb{j}_i(\vb{k},\omega)} \vdot \var{ \vb{E}_Z (\vb{-k}, -\omega)} \rangle ,
\end{multline}
where $\var{\vb{j}_i}$ is the ion current associated with the induced density change, $\vb{k}\vdot\var{\vb{j}_i} = e \omega \var{n_i}$, and $\var{\vb{E}_Z}$ is the change in electric field, $\var{\vb{E}_Z} = \imath \vb{k} \var{ \phi_Z }$.
With these two relations, this becomes
\begin{multline}
	\dv{E_i}{t}_{e \rightarrow i} = - n_i \int \frac{\dd{\vb{k}}}{(2\pi)^3} \int \frac{\dd{\omega}}{2\pi} \omega \left( \frac{4 \pi e^2}{k^2} \right)^2 \\
	\times \Im\left[ \chi_{ii}(\vb{k}, \omega) \langle \var{ Z(\omega) } \var{ Z(-\omega) } \rangle \right] .
\end{multline}
Due to detailed balance of the ionization and recombination rates, the charge state fluctuations will obey a fluctuation-dissipation theorem \cite{Lakatos-Lindenberg1972}
\begin{equation}
    \langle \var{ Z(\omega) } \var{ Z(-\omega) } \rangle = -\hbar \coth(\frac{\hbar \omega}{2 T_e}) \Im \chi_{ZZ}(\omega) ,
\end{equation}
where $\chi_{ZZ}$ is a response function for the charge state.
We can use this to write the energy transfer as
\begin{multline}
    \dv{E_i}{t}_{e \rightarrow i} = n_i \int \frac{\dd{\vb{k}}}{(2\pi)^3} \int \frac{\dd{\omega}}{2\pi} \hbar \omega \left( \frac{4 \pi e^2}{k^2} \right)^2 \\
    \times \Im \chi_{ii}(\vb{k}, \omega) \Im \chi_{ZZ}(\omega) \coth(\frac{\hbar \omega}{2 T_e}) .
\end{multline}

In equilibrium, there should be no net energy transfer.  
There must therefore be an inverse process, which will be in detailed balance when the electron and ion temperatures are equal.  
The inverse process can be understood by considering the influence of the ion microfields on the bound states.
Spontaneous ion density fluctuations lead to time-varying microfields which modify the ionization potential
\begin{equation}
	\label{eq:delta_I}
    \var{I}(\omega) = \int \frac{\dd{\vb{k}}}{(2\pi)^3} \frac{4 \pi e^2}{k^2} \var{ n_i }(\vb{k},\omega).
\end{equation}
This perturbs the probabilities for ionization and recombination.  
When the ionization potential is reduced, ionization becomes more likely.  
Conversely, when the ionization potential is increased, recombination becomes more likely.  This leads to a net transfer of energy to the electrons, as they gain more energy during recombination than is lost during ionization.  
Again using linear response theory, the average change in the charge state is
\begin{equation}
    \var{Z}(\omega) = \chi_{ZZ}(\omega) \var{I}(\omega).
\end{equation}
The rate of energy change is then given by
\begin{align}
    \dv{E_i}{t}_{i \rightarrow e} &= - n_i \int \frac{\dd{\omega}}{2\pi} \imath \omega \langle \var{Z(\omega)} \var{ I( -\omega) } \rangle \\
    &= n_i  \int \frac{\dd{\omega}}{2\pi} \omega \Im\left[ \chi_{ZZ}(\omega) \langle \var{I}(\omega) \var{I}(-\omega) \rangle \right]
\end{align}
Now inserting Eq.~\ref{eq:delta_I}, together with the fluctuation-dissipation relation for the ions
\begin{equation}
        \langle \var{ n_i(\vb{k},\omega) } \var{ n_i(\vb{-k},-\omega) } \rangle = -\hbar \coth(\frac{\hbar \omega}{2 T_i}) \Im \chi_{ii}(\vb{k},\omega) ,
\end{equation}
the rate of energy transfer from the ions to the electron can be written as 
\begin{multline}
   \dv{E_i}{t}_{i \rightarrow e} = -n_i \int \frac{\dd{\vb{k}}}{(2\pi)^3} \int \frac{\dd{\omega}}{2\pi} \hbar \omega \left( \frac{4 \pi e^2}{k^2} \right)^2 \\
   \times \Im \chi_{ii}(\vb{k}, \omega) \Im \chi_{ZZ}(\omega) \coth(\frac{\hbar \omega}{2 T_i}) .
\end{multline}

The net exchange rate, including both the forward and reverse processes, is then given by
\begin{multline}
    \dv{E_i}{t} = n_i \int \frac{\dd{\vb{k}}}{(2\pi)^3} \int \frac{\dd{\omega}}{2\pi} 		\hbar \omega \left( \frac{4 \pi e^2}{k^2} \right)^2 \\
    \times \Im \chi_{ii}(\vb{k}, \omega) \Im \chi_{ZZ}(\omega) \\
    \times \left[ \coth(\frac{\hbar \omega}{2 T_e}) - \coth(\frac{\hbar \omega}{2 T_i}) \right] .
\end{multline}
This expression closely resembles the exchange rate for direct collisions, except that the electron fluctuation spectrum has been replaced by the spectrum of charge state fluctuations, $\Im \chi_{ZZ}$.

\begin{figure}
    \centering
    \includegraphics[width=\linewidth]{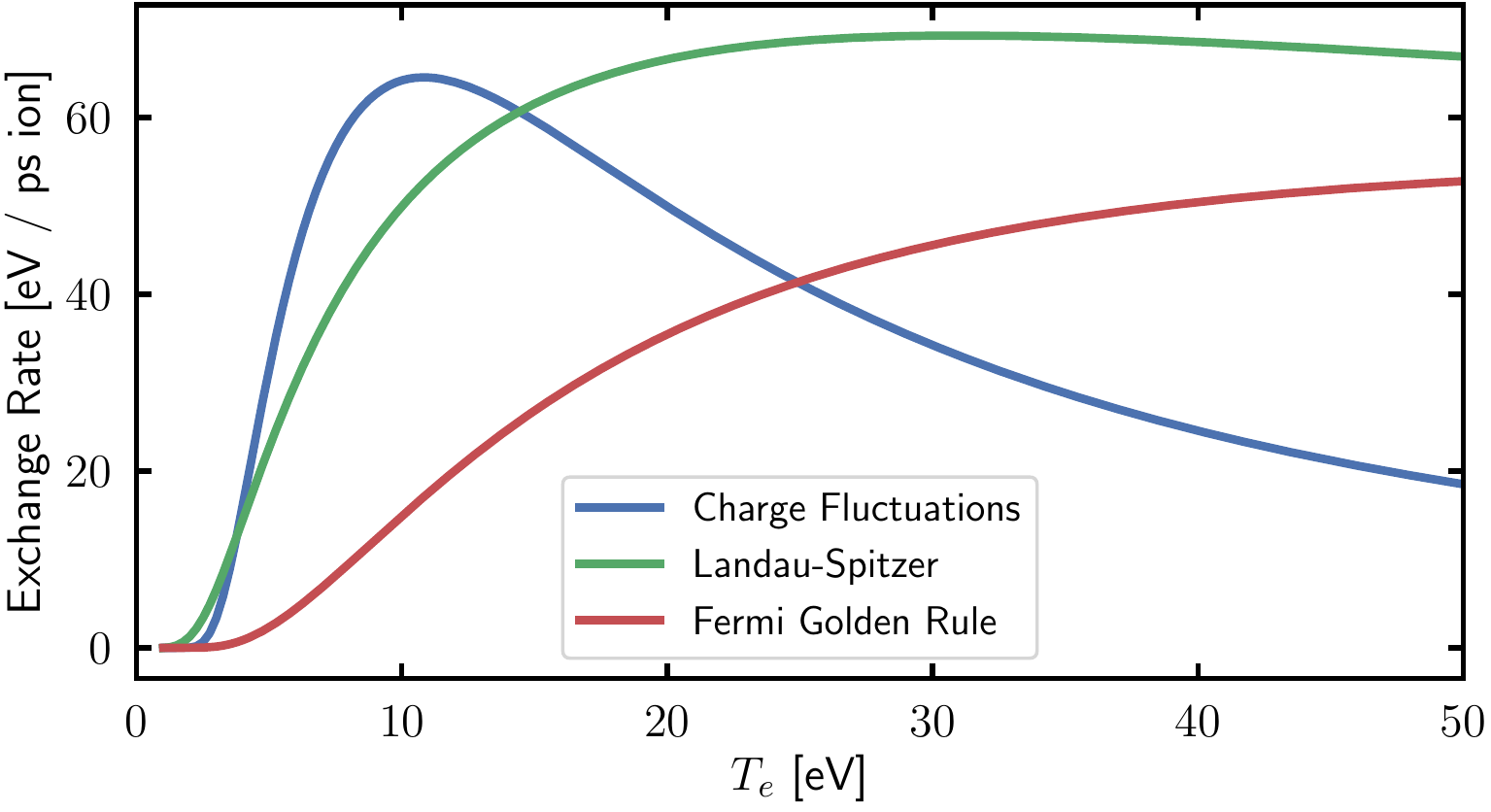}
    \caption{ \label{fig:h_rates}
    Electron-ion energy transfer rate in hydrogen at liquid density, $\rho = \SI{0.07}{g.cm^{-3}}$ and with an ion temperature of $T_i=\SI{1}{eV}$.  The rate due to charge state fluctuations is compared to the rate from direct interactions predicted by the Landau-Spitzer \cite{Gericke2002} and Fermi Golden Rule \cite{Dharma-wardana1998} approaches.}
\end{figure}

In order to carry out calculations of this new exchange rate, we have modelled the charge state fluctuations as a random walk, with the probability of jumps between charge states determined from the collisional ionization and recombination rates \footnote{Please see Supplemental Material, which includes Refs.~\cite{Grebenkov2018,Chung2005,Lotz1968,Schlanges1988,Atzeni2004,Bornath1993,Ciricosta2012,Hoarty2013,Stewart1966,Ecker1963,Murillo1998}, for further details of the random walk model.}.
The effect of ionization potential depression (IPD) on the ionization rates has been included using the Stewart-Pyatt model \cite{Stewart1966}, although calculations with other widely-used IPD models \cite{Ecker1963,Murillo1998} showed few significant differences.
Electron degeneracy has also been included through the use of an interpolative form for the chemical potential \cite{Atzeni2004}.

The random walk yields a fluctuation spectrum of the form
\begin{equation}
	\label{eq:corr_spectrum}
	\langle \var{ Z(\omega) } \var{ Z(-\omega) } \rangle
	= \frac{1}{\pi} \sum_i \frac{A_i \tau_i}{1+\omega^2\tau_i^2} .
\end{equation}
The number of components in the spectrum is in principle equal to the atomic number of the ion.
In the case of hydrogen only a single component is present, for which the correlation amplitude $A$, and decay time $\tau$, can be expressed analytically in terms of the ionization and recombination coefficients.  

Calculated energy transfer rates in liquid-density hydrogen, which can be measured in pump-probe XFEL experiments \cite{Zastrau2014}, are shown in Fig.~\ref{fig:h_rates}.
As our focus in this work is on the charge fluctuations, the ion density response has been calculated using the random phase approximation \cite{Bohm1953} throughout.
For temperatures up to the ionization potential, the energy transfer due to charge state fluctuations is comparable to the Landau-Spitzer predictions for direct interactions.
At higher temperatures, the charge fluctuation mechanism becomes less significant as the plasma increasingly tends towards complete ionization.

\begin{figure}
    \centering
    \includegraphics[width=\linewidth]{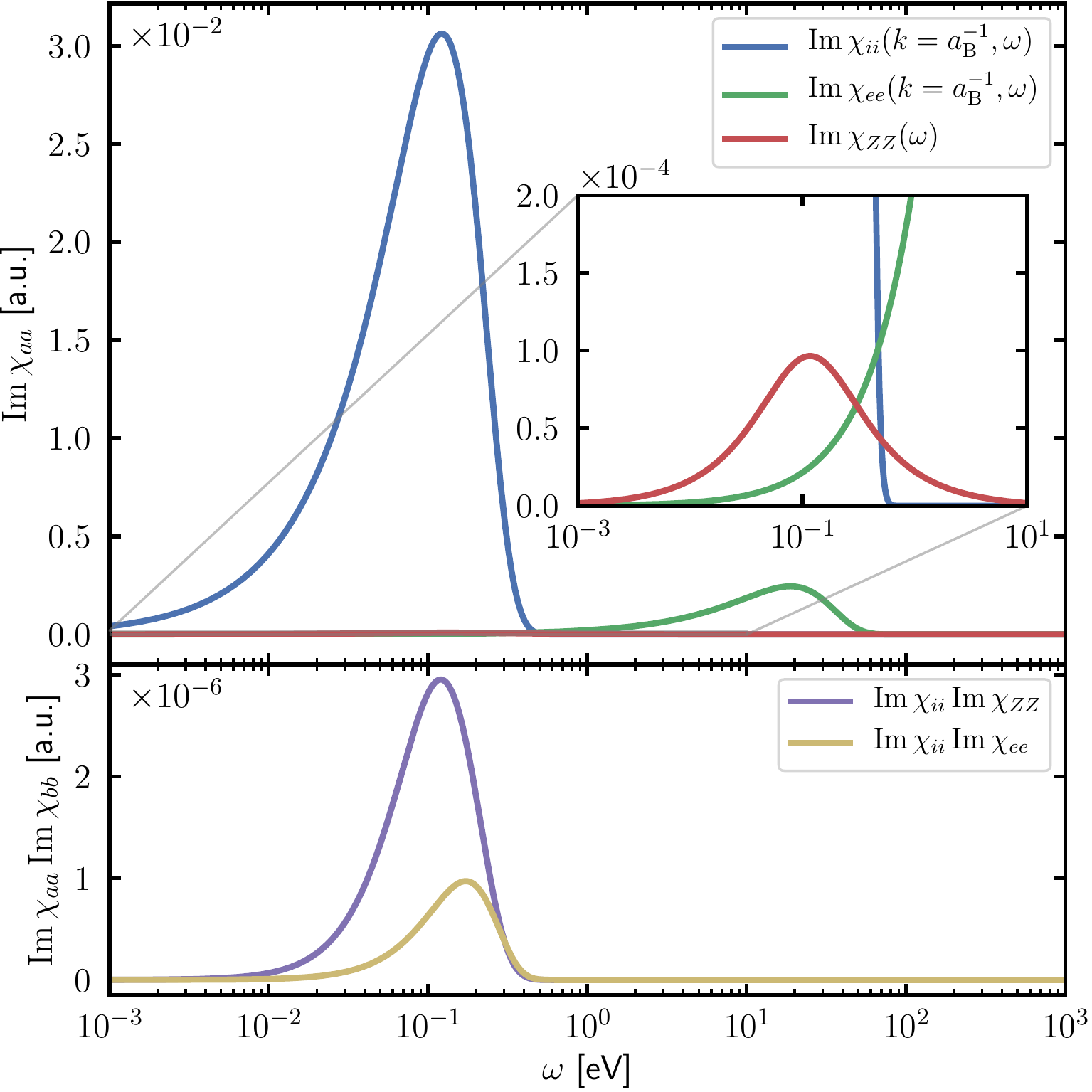}
    \caption{\label{fig:resp_funcs}
    Upper: Response functions for the electrons, ions and charge state in hydrogen with $T_i=\SI{1}{eV}$, $T_e=\SI{10}{eV}$ and $n_e=\SI{1e22}{cm^{-3}}$.  The inset shows the charge state response function on a larger scale.
    Lower: Products of the ion response function with the electron and charge state response functions.}
\end{figure}

Although this is an indirect process, in the sense that the electrons and ions interact via a third body (an ion with fluctuating charge), it is efficient at transferring energy.
This can be understod by considering the spectrum of charge state fluctuations alongside the electron and ion modes, as shown in Fig.~\ref{fig:resp_funcs}.
Direct energy exchange is a slow process due to inefficient coupling between fast electron modes and slow ion modes; there is very little overlap between ion and electron response functions.
In contrast, the charge state fluctuations overlap more strongly with the ion modes.
This leads to an efficient energy transfer, even though the charge state fluctuations are relatively weak in comparison to the density fluctuations.

\begin{figure}
    \centering
    \includegraphics[width=\linewidth]{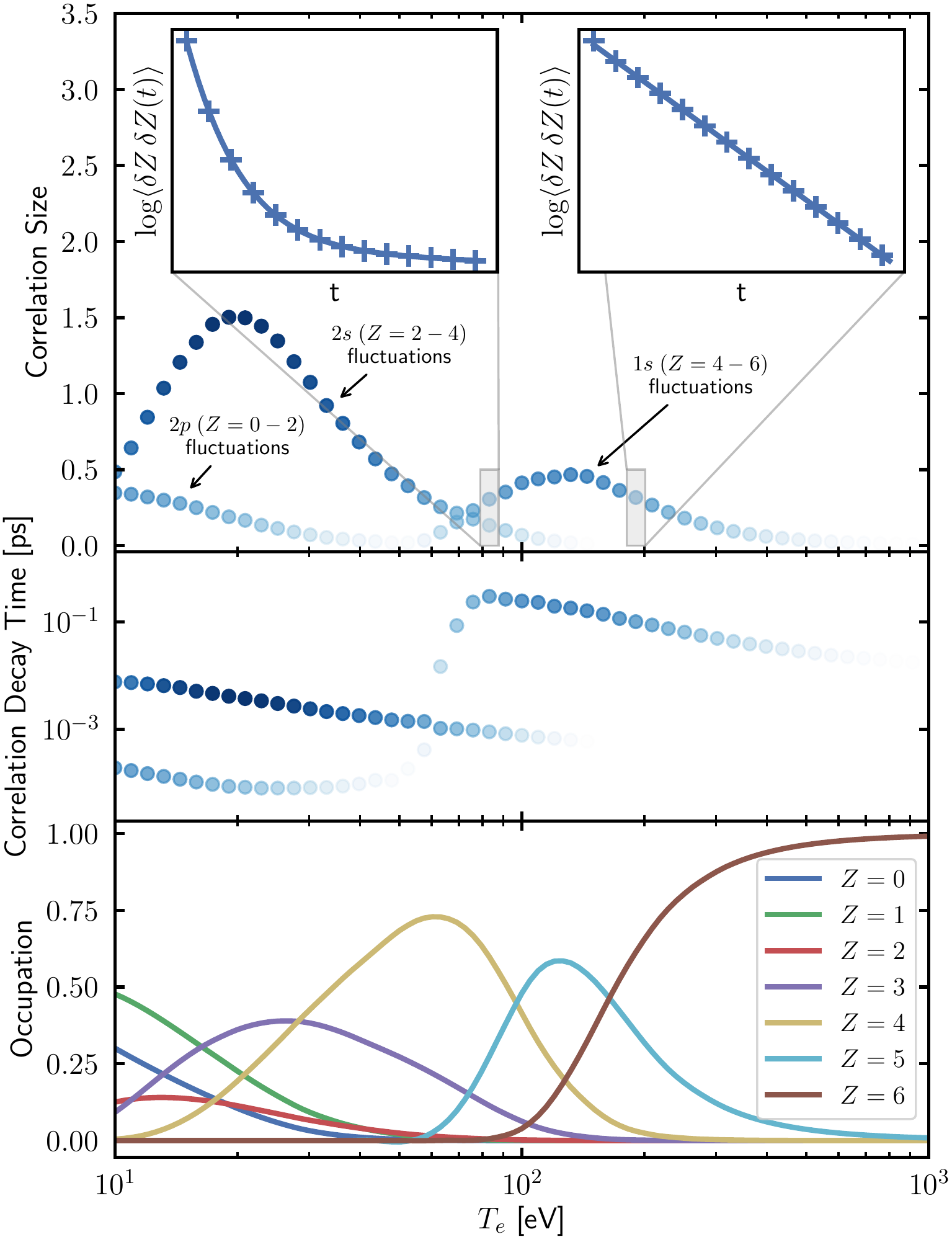}
    \caption{\label{fig:monte_c}
    Upper and Middle: Fluctuation size and decay times (i.e., the $A_i$ and $\tau_i$ of Eq.~\ref{eq:corr_spectrum}) in solid-density carbon obtained from Monte Carlo simulations.  The color scale denotes the correlation size, so that larger fluctuations are the most prominent.  The insets show the fit to selected correlation functions, as used to obtain the size and decay rates.  Lower: The corresponding charge state occupations. }
\end{figure}

\begin{figure}
    \centering
    \includegraphics[width=\linewidth]{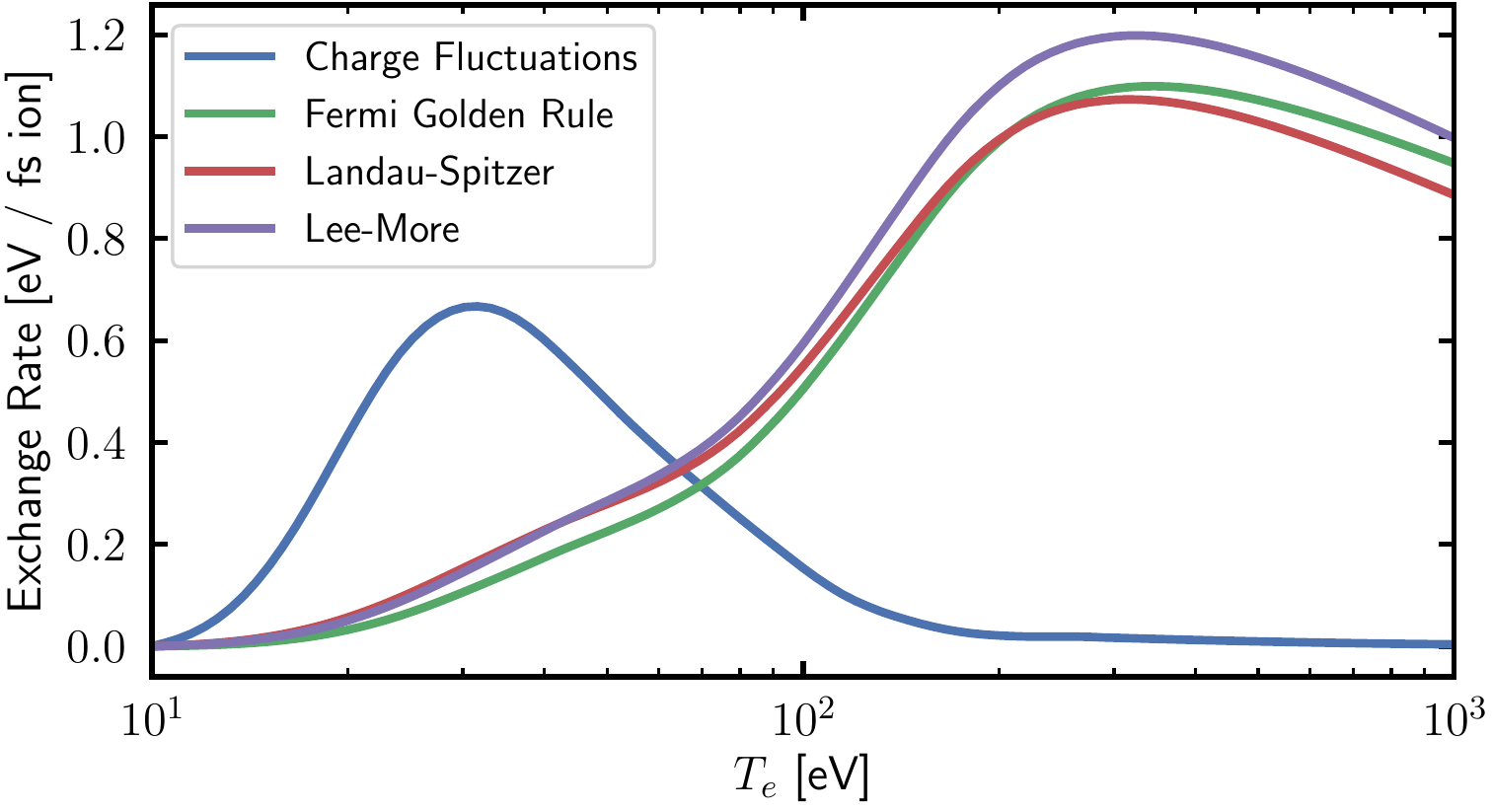}
    \caption{\label{fig:solid_c}
    Electron-ion energy transfer rates in solid-density carbon, comparing the rate due to charge fluctuations with different models for direct electron-ion collisions \cite{Dharma-wardana1998, Gericke2002, Lee1984}.  The ion temperature is $T_i=\SI{10}{eV}$. }
\end{figure}

For elements with higher atomic numbers, obtaining the charge state fluctuation spectrum analytically becomes intractable.
Instead, the random walk can be simulated numerically using a Monte Carlo approach.
We can then fit the correlation function to obtain the $A_i$ and $\tau_i$ of Eq.~\ref{eq:corr_spectrum}.
Results for solid-density carbon, where measured energy transfer rates significantly exceeded previous predictions \cite{Hau-Riege2012}, are shown in Fig.~\ref{fig:monte_c}.
Although the spectrum at each temperature could comprise up to six components in principle, we find that no more than two are required to obtain a good fit for the conditions studied here (i.e., only two terms contribute significantly to the sum in Eq.~\ref{eq:corr_spectrum}).
This is representative of the fact that, at a given temperature and density, only a subset of charge states show significant occupation.
Comparing the components of the fluctuation spectrum with the charge state occupations, we can broadly identify three contributions, corresponding to fluctuating ionization of the $1s\;(Z=4-6)$, $2s\;(Z=2-4)$ and $2p\;(Z=0-2)$ electrons.

The corresponding energy transfer rates for solid-density carbon are shown in Fig.~\ref{fig:solid_c}.
As in the hydrogen case, predicted energy transfer rates from charge fluctuations are comparable with direct electron-ion interactions.
Notably, the peak transfer rate occurs at lower temperatures, where the ionization of the $2s$ electrons dominates the fluctuations.
At higher temperatures, fluctuations in the ionization of the $1s$ electrons occur, but the slower ionization rate for these states means that these fluctuations do not couple as strongly to the ion modes.
This suggests that the range of temperatures where this additional mechanism is significant might be restricted, even where more electron shells are present.

We can also calculate energy transfer rates for the conditions probed in Ref.~\cite{Hau-Riege2012} more specifically.
There, the ion temperature increases from $T_i = \SI{1}{\eV}$ to $T_i = \SI{5}{\eV}$ and the electron temperature is bounded by $T_i \leq T_e \leq \SI{10}{\eV}$.
This range of electron temperatures leads to energy exchange rates of $\SIrange[range-phrase=-, range-units=single]{0.03}{0.27}{\eV \per \fs}$ due to charge state fluctuations, compared to the measured value of $\SI{0.17}{\eV \per \fs}$ given in Ref.~\cite{Hau-Riege2012}.
It is feasible, therefore, that charge state fluctuations are significant in this case, although a more precise conclusion is limited by the experimental uncertainty in electron temperature.

\begin{figure}
    \centering
    \includegraphics[width=\linewidth]{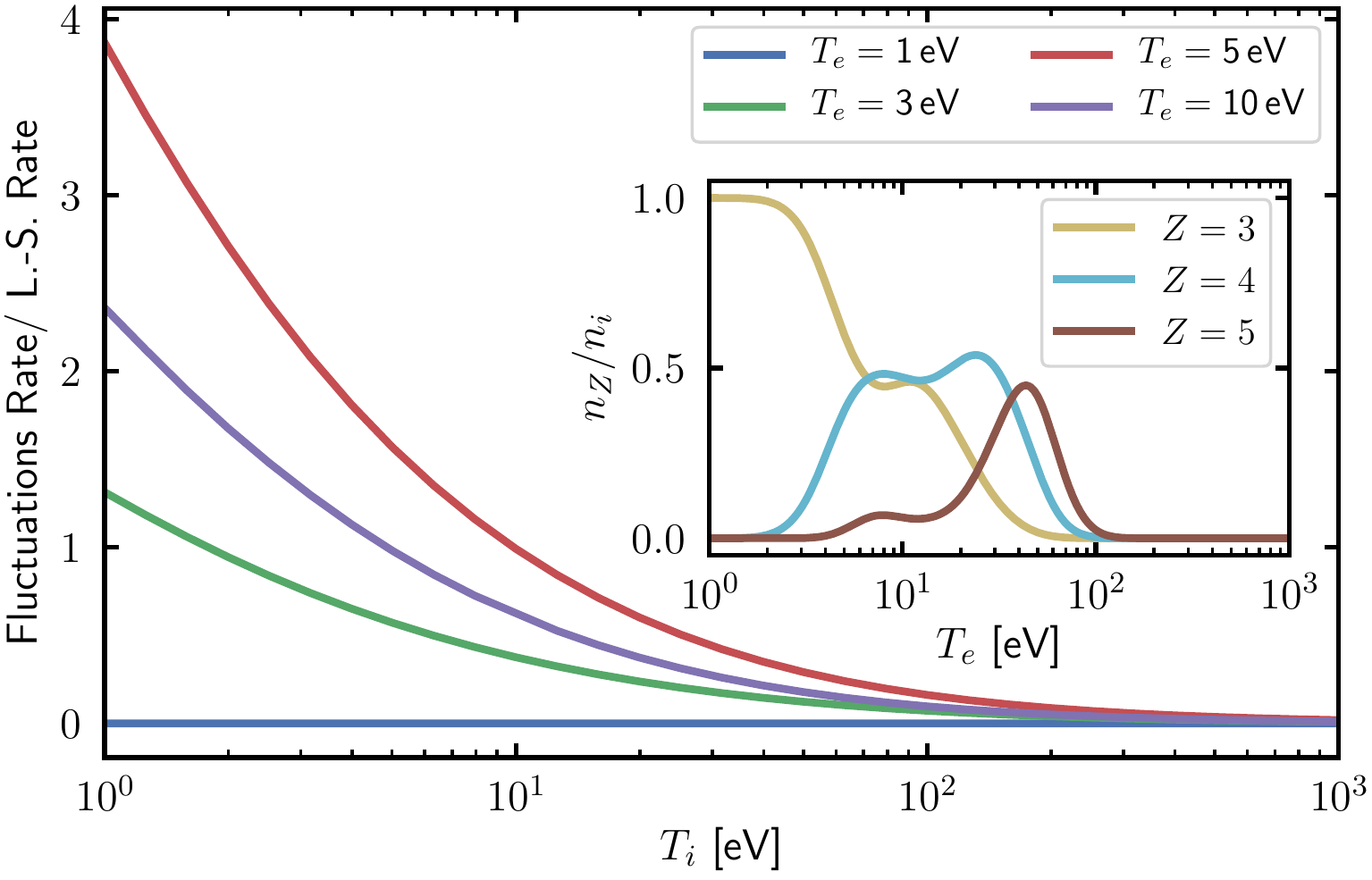}
    \caption{\label{fig:i_to_e}
    Ratio of the charge fluctuation rate to the Landau-Spitzer rate in solid-density aluminium with hot ions and cooler electrons.  The inset shows the occupation of the charge states as a function of electron temperature. This calculation uses the Debye IPD model \cite{Murillo1998}, which more accurately predicts the pressure ionized M-shell expected in solid-density aluminium.  }
\end{figure}

In metallic systems, the outer shell electrons are pressure ionized at low electron temperatures.
This precludes fluctuations in the ionization of these shells and can strongly suppress this energy exchange mechanism.  
This is particularly true for energy transfer from hot ions to cooler electrons, as in this situation the electrons do not have sufficient energy to ionize remaining bound electrons.
This is shown in the case of solid-density aluminium in Fig.~\ref{fig:i_to_e}.  
As such, we might expect charge state fluctuations to make a limited contribution to energy transfer in shock-heated systems.  This could explain the lower equilibration rates observed in shock-heated metals \cite{Ng1995,Riley2000} compared to XFEL-heated carbon \cite{Hau-Riege2012}.

In summary, energy transfer through charge state fluctuations can be a significant effect in partially ionized dense plasmas.
Moreover, this mechanism exhibits a non-trivial dependence on the temperature, and on the structure of the material through effects such as pressure ionization.
This could contribute to our understanding of variability in experimental measurements of electron-ion equilibration.

Although this work has focussed on charge state fluctuations due to collisional processes, fluctuations can also be driven by radiative processes when sufficiently intense radiation fields are present.
Under these conditions, the mechanism presented here would directly couple the radiation and ion temperatures, which are otherwise only coupled via the electrons.

\begin{acknowledgments}
This project has received funding from the European Research Council (ERC) under the European Union's Horizon 2020 research and innovation programme (grant agreement no. 682399).
\end{acknowledgments}

\bibliography{../fluctuations2020}

\end{document}


\title{Supplemental Material to: Temperature Equilibration Due to Charge State Fluctuations in Dense Plasmas}

\newcommand{\Phys}
{Plasma Physics Group, Blackett Laboratory, Imperial College London, London, SW7 2AZ, UK}

\author{R.~A.~Baggott}
\email{r.baggott@imperial.ac.uk}
\affiliation{\Phys}

\author{S.~J.~Rose} 
\affiliation{\Phys}

\author{S.~P.~D.~Mangles} 
\affiliation{\Phys}

\maketitle

\section{Random Walk Model}
A random walk process is characterized by its transition matrix.
We can construct a transition matrix for the random walk between ionization stages by considering the properties of collisional ionization and recombination.
First, we assume only single ionization and recombination processes; processes such as double impact ionization, which have much lower cross-sections, are neglected, but could be included in principle.
The charge state can therefore only change by $\pm 1$ at each event.
Since the ionization and recombination coefficients are constant, the event times are drawn from an exponential distribution.
The ionization and recombination rates between charge states $Z$ and $Z+1$ depend on the ionization potential separating the two states.
The random walk is therefore heterogeneous; the event distributions vary with the charge state.
The transition matrix can be most conveniently handled in Laplace space, where it has
a tridiagonal form:
\begin{equation}
     Q_{ij}(s) = 
    \begin{cases}
        (1+\mathcal{T}_{\alpha i}s)^{-1} & \text{if } j=i+1, \\
        (1+\mathcal{T}_{\beta j}s)^{-1}  & \text{if } j=i-1, \\
        0 & \text{otherwise}.
    \end{cases}
\end{equation}
The characteristic times for ionization and recombination, $\mathcal{T}_{\alpha i}$ and $\mathcal{T}_{\beta i}$, are given by $\mathcal{T}_{\alpha i} = 1/n_e\alpha_i$ and $\mathcal{T}_{\beta i} = 1/n_e^2\beta_i$, where $\alpha_i$ and $\beta_i$ are the ionization and recombination coefficients.

Once the transition matrix has been constructed, the properties of the random walk can be determined by obtaining the propagator.
The propagator gives the probability to find the system in state $x$ at time $t$ given that it was in state $x_0$ at $t=0$.
For the heterogeneous random walk, the propagator is given by \cite{Grebenkov2018}
\begin{equation}
        P_{x_0x} \left(s\right) = \frac{1 - \sum_{x'}Q_{xx'}(s)}{s}\left[ \left(I - Q(s) \right)^{-1} \right]_{x_0x} ,
\end{equation}
where $I$ is the identity matrix.
The correlation function for charge fluctuations can be expressed in terms of the propagator
\begin{equation}
	\label{eq:corr_func}
        \left<\delta Z(t) \delta Z(0) \right> = \sum_{x,x_0} Z_x Z_{x_0} P_{x_0 x} p^\text{st}_{x_0} - \left(\sum_{x_0} Z_{x_0}p^\text{st}_{x_0}\right)^2 ,
\end{equation}
where $p^\text{st}$ are the occupations of the charge states in steady state.
The decay properties of the correlations are therefore determined by the propagator.

In general, $P_{x_0x}(s)$ is a ratio of two polynomials in $s$.  
We can therefore carry out the inverse Laplace transform, using a partial fraction expansion, to show that $P_{x_0x}(t)$ will be a sum over decaying exponentials, with the decay times $\tau_i$ given by the roots of the denominator, i.e., $\det[I-Q(s)]=0$.
Substituting this into Eq.~\ref{eq:corr_func}, we obtain the general form of the correlation function
\begin{equation}
	\label{eq:corr_func_exp}
	\left<\delta Z(t) \delta Z(0) \right> = \sum_i A_i \exp\left({-\frac{\abs{t}}{\tau_i}}\right).
\end{equation}

For hydrogen, with only two ionization stages, we can find the roots analytically to obtain
\begin{equation}
        \left<\var{ Z(t)} \var{ Z(0)} \right> = \frac{\mathcal{T}_\alpha \mathcal{T}_\beta}{(\mathcal{T}_\alpha + \mathcal{T}_\beta)^2} \exp\left(-\frac{\mathcal{T}_\alpha + \mathcal{T}_\beta}{\mathcal{T}_\alpha \mathcal{T}_\beta}\abs{t}\right) .
\end{equation}
For the case where $\mathcal{T}_\beta >> \mathcal{T}_\alpha$ (or $\mathcal{T}_\alpha >> \mathcal{T}_\beta$) the system will be fully ionized (un-ionized) with only small fluctuations.  
In this case we can see that the fluctuation decay rate reduces to the ionization (recombination) rate.

For heavier elements, finding the roots analytically becomes intractable.  For these elements, we carry out Monte Carlo simulations of the random walk and calculate the correlation function numerically.  We then fit the known exponential form of the correlations to the numerical result, in order to obtain decay rates and amplitudes.  Hypothesizing that only a subset of the exponentials will contribute strongly, we initially fit only a single exponential and add subsequent exponentials only when the quality of the fit, determined by the $R^2$ value, can be improved.

The energy transfer calculation uses the charge fluctuation spectrum in frequency space,
\begin{equation}
	\label{eq:corr_spectrum}
	\langle \var{ Z(\omega) } \var{ Z(-\omega) } \rangle
	= \frac{1}{\pi} \sum_i \frac{A_i \tau_i}{1+\omega^2\tau_i^2} ,
\end{equation}
which is obtained through a Fourier transform of Eq.~\ref{eq:corr_func_exp}.
The correlation amplitudes and decay rates, obtained either analytically or through the Monte Carlo approach, are substituted into this expression.

\section{Rate Coefficients}

The collisional ionization and recombination rates depend principally on the distribution function of the free electrons and on the collisional cross-section.  
A number of semi-empirical cross-sections can be applied with reasonable success and remain widely used \cite{Chung2005}.  
One such cross-section is due to Lotz \cite{Lotz1968}, which we adopt in this work.
With a Maxwellian electron distribution, this leads to an ionization rate \cite{Schlanges1988}
\begin{equation}
    \alpha_i^\text{ideal} = \frac{10 \pi a_\text{B}^2}{(2 \pi m_e k_B T_e)^{1/2}} \frac{\si{Ry^2}}{I_i} \Ei \left(-\frac{I_i}{k_B T_e}\right) ,
\end{equation}
where $I_i$ is the ionization potential and $\si{Ry}$ is the Rydberg energy.
The recombination rate can then be determined from detailed balance
\begin{equation}
    \beta_i = \alpha_i \frac{n_i}{n_e n_{i+1}} \exp \left[ \flatfrac{\left( \mu_e + \mu_{i+1} - \mu_i\right)}{k_B T_e} \right] .
\end{equation}
The effect of electron degeneracy can be included through an appropriate form for the electron chemical potential, $\mu_e$.
We have used the interpolation due to Ichimaru, which incorporates the classical and quantum limits and is accurate to within $0.3\%$ in the intermediate regime \cite{Atzeni2004}.

\begin{figure}
    \centering
    \includegraphics[width=\linewidth]{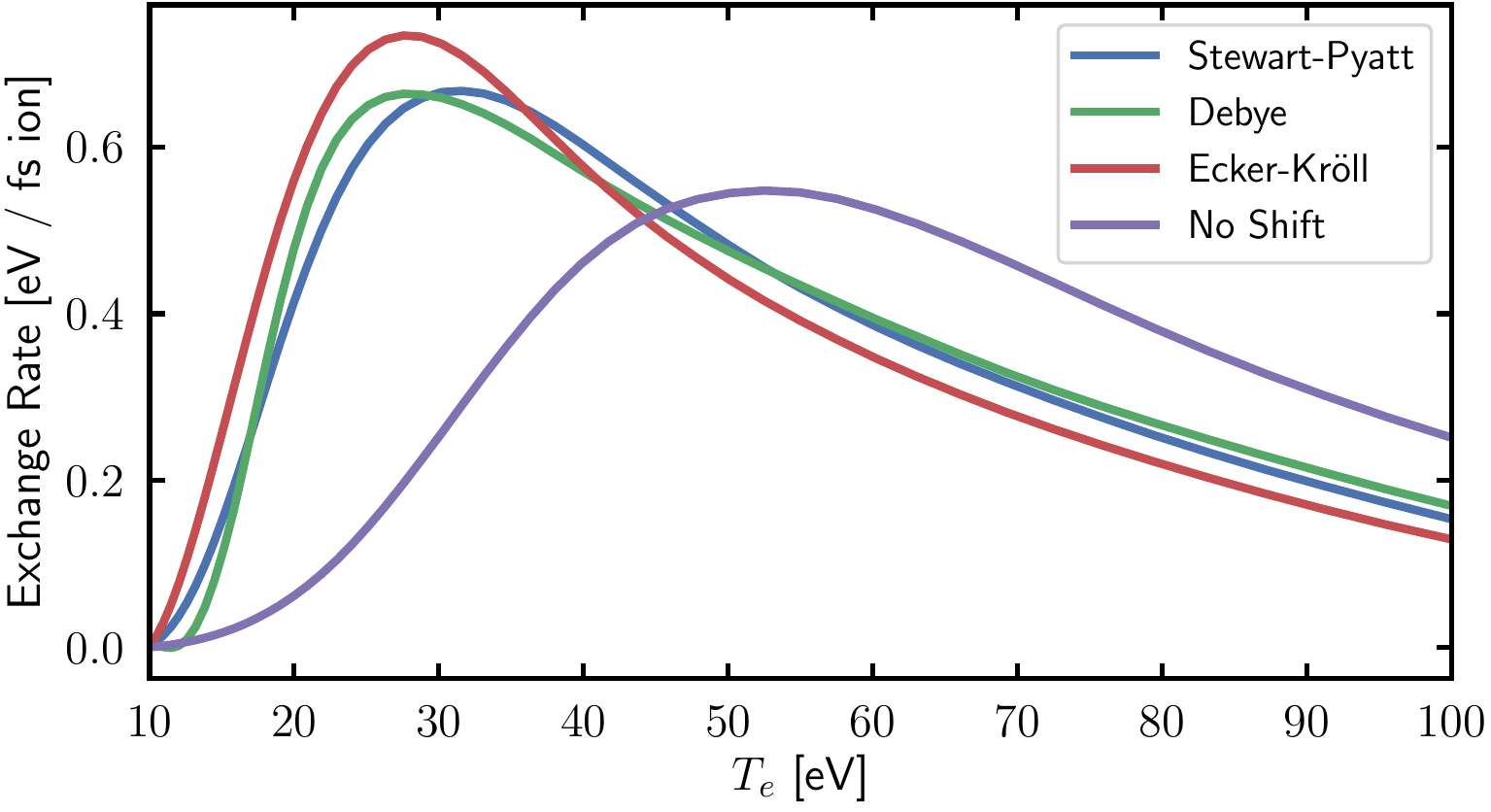}
    \caption{\label{fig:ipd}
    Comparison of electron-ion energy transfer rates in solid-density carbon with $T_i=\SI{10}{eV}$, calculated with different IPD models \cite{Stewart1966,Ecker1963,Murillo1998}.  }
\end{figure}

In dense plasmas, the detailed balance, and hence the level of ionization, is modified by the presence of ionization potential depression (IPD).
To a first approximation, this modification manifests as an increase in the ionization rate, while the recombination rate remains relatively unchanged \cite{Bornath1993}.
This effect has therefore been included by modifying the ionization rate according to 
\begin{equation}
    \alpha_i = \alpha_i^\text{ideal} \exp \left[ \frac{\Delta I_i}{k_B T_e} \right] .
\end{equation}
A number of models exist for the IPD, $\Delta I_i$, and the accuracy of the most common models is a topic of ongoing debate \cite{Ciricosta2012,Hoarty2013}.
While the IPD can modify the charge fluctuation spectrum by increasing the ionization rate, the fact that the recombination rate remains unchanged means that this is necessarily accompanied by a significant change in ionization.
There is therefore limited scope for the IPD to modify the charge fluctuation spectrum; changing the IPD enough to materially alter the fluctuation spectrum would lead to an anomalous level of ionization.
This is supported by our calculations with various IPD models, as shown in Fig.~\ref{fig:ipd}.
Although it is important to include IPD in some form, the properties of the charge fluctuations, and therefore the energy exchange rate, were not sensitive to the choice of model.

\bibliography{../fluctuations2020}